\documentclass[10pt,conference]{IEEEtran}

\usepackage{iftex}
\ifpdftex
\usepackage[utf8]{inputenc}
\usepackage[T1]{fontenc}
\usepackage{booktabs}
\fi

\usepackage{hyperref}

\usepackage{graphicx}
\usepackage{float}
\usepackage{url}
\usepackage[para]{footmisc}
\usepackage[inline]{enumitem}

\usepackage{titlesec}
\titlespacing*{\section}{0pt}{*2}{*1}
\titlespacing*{\subsection}{0pt}{*2}{*1}
\titlespacing*{\subsubsection}{0pt}{*2}{*1}

\begin{document}

\title{Towards Continuous Experiment-driven MLOps}
\newcommand{\projectName}{ExtremeXP}


%
\author{\IEEEauthorblockN{Keerthiga Rajenthiram\IEEEauthorrefmark{1},
Milad Abdullah\IEEEauthorrefmark{2},
Ilias Gerostathopoulos\IEEEauthorrefmark{1}, 
Petr Hnětynka\IEEEauthorrefmark{2},
Tomáš Bureš\IEEEauthorrefmark{2}},
Gerard Pons\IEEEauthorrefmark{3},
Besim Bilalli\IEEEauthorrefmark{3} and
Anna Queralt\IEEEauthorrefmark{3}
\IEEEauthorblockA{\IEEEauthorrefmark{1}Vrije Universiteit Amsterdam, The Netherlands  \{k.rajenthiram, i.g.gerostathopoulos\}@vu.nl}
\IEEEauthorblockA{\IEEEauthorrefmark{2}Charles University, Prague, Czech Republic \{milad.abdullah, petr.hnetynka, tomas.bures\}@matfyz.cuni.cz}
\IEEEauthorblockA{\IEEEauthorrefmark{3}Universitat Polit\`ecnica de Catalunya, Barcelona, Spain \{gerard.pons.recasens, besim.bilalli, anna.queralt\}@upc.edu
}}

\maketitle

\begin{abstract}
Despite advancements in MLOps and AutoML, ML development still remains challenging for data scientists. 
First, there is poor support for and limited control over optimizing and evolving ML models. 
Second, there is lack of efficient mechanisms for continuous evolution of ML models which would leverage the knowledge gained in previous optimizations of the same or different models. 
We propose an experiment-driven MLOps approach which tackles these problems. 
Our approach relies on the concept of an experiment, which embodies a fully controllable optimization process. 
It introduces full traceability and repeatability to the optimization process, allows humans to be in full control of it, and enables continuous improvement of the ML system. 
Importantly, it also establishes knowledge, which is carried over and built across a series of experiments and allows for improving the efficiency of experimentation over time. 
We demonstrate our approach through its realization and application in the \projectName{}\footnote{https://extremexp.eu/} project (Horizon Europe).
\end{abstract}


\section{Introduction}

Although ML has transformed business and society in many areas (like crisis management, transportation, agriculture, and biology), the application of ML and, in particular, the development, maintenance, and evolution of ML models and corresponding ML-enabled systems are still largely considered an art---rather than an engineering discipline---informed by the experience and expertise of data scientists~\cite{jung-lin_lee_human---loop_2019, xin_whither_2021}. 

So far, two main paradigms have been proposed to streamline the development of ML-enabled systems: MLOps and AutoML. 
MLOps focuses on the process of developing and operating ML-enabled systems, bridging the gap between training ML models and deploying and maintaining them~\cite{kreuzberger_machine_2023}.
AutoML instead focuses on increasing automation in ML optimization (model hyperparameter optimization, but also tuning of, e.g., data cleaning and feature selection) and by that, making ML more accessible to less technical users~\cite{xin_whither_2021, xanthopoulos_putting_2020}. 
AutoML is usually seen as part of MLOps.

Despite the efficiency gains obtained by combining MLOps practices and AutoML frameworks, we argue that data scientists working with complex ML workflows are still facing important challenges in their daily work. 

First, data scientists \textit{spend a large amount of time and resources} in optimizing and evolving their ML workflows~\cite{rajenthiram_optimizing_2024}. 
Such evolution is by nature an iterative process where a data scientist introduces a change to a workflow, executes the workflow, measures the effect of the change, and decides which change to do next. 
It is therefore inevitable to spend time and resources in (i) recomputing workflows, (ii) analyzing the results and deciding on the next trials.  
While approaches do exist to make (i) more efficient (e.g., by re-computing only the affected parts of the workflow~\cite{xin_helix_2018}), support for (ii) is almost absent in today's MLOps/AutoML frameworks. 
Having executed a number of optimization and evolution steps, such frameworks should ideally guide the user by extracting knowledge and patterns from past evolution steps---of both the same and of similar workflows---and proposing meaningful next steps.  

Second, data scientists \textit{lack fine-grained control over the optimization} of their workflows. 
Broadly, when it comes to optimizing the performance of a workflow, a data scientist can choose one of two alternatives: either manually run specific configurations of the workflow or rely on an AutoML framework to perform auto-tuning. 
Such frameworks, however, do not allow the user to interfere with or guide the optimization process.
Ideally, they should do so by some sort of user interaction ``checkpoints'' where the system executes a number of workflows, shows the users the results obtained so far, and allows them to change the workflows that are scheduled for execution based on the intermediate results. 
Although some frameworks allow for coarse-grained interaction by, e.g., stopping the execution of a step in the optimization process (e.g., MLFlow), the above-described fine-grained interaction---although important in taking advantage of the latent knowledge and expertise of data scientists~\cite{xin_whither_2021}---is largely missing. 

Third, data scientists \textit{lack efficient mechanisms for continuous evolution} of ML models based on production data.  
In a typical case, once a model is optimized/evolved to a sufficient degree, it is deployed in production, where its performance and the data inputs and outputs may be monitored for detecting drifts and regressions. 
Runtime data may be used then in offline (or, in advanced cases, online) re-training of the ML model. 
Ideally, such re-training however should consider the knowledge gathered in the previous optimization and evolution steps, including knowledge about optimizations that did not yield expected performance gains. This is crucial in order to bring time- and resource-efficiency in the re-training. 
With the current MLOps and AutoML frameworks and practices, it is still difficult to continuously evolve and optimize a model \textit{after deployment} in an efficient way, as the past knowledge cannot be easily taken into account.

In response to these challenges, we propose to rethink ML optimization and evolution as a series of experiments performed by data scientists. 
This is in line with the vision of Bosch and Olsson for building autonomous systems which suggests that we need a ``\textit{balance where R\&D teams build part of the functionality and set guardrails, and where smart systems} (in our case, self-optimizing ML workflows) \textit{experiment and adjust their responses and behaviors autonomously}''~\cite{BoschEvolution2026}.

Our experiment-driven MLOps approach differs from regular practices in three important ways. 
First, it learns from the past by assuming full traceability of data, results, and code, full repeatability of experiments, and having an explicit notion of the cost of re-computing workflows.  
Second, it allows humans to be in control of the optimization process, not to be just initiators or observers of it. 
Third, it allows ML systems to continuously improve by collecting more evidence, even in production environments, about their actual performance. 


The main contribution of this paper is to (i) motivate and position experiment-driven MLOps (developed within the \projectName{} project) and (ii) describe its main concepts.

\section{Background and Related Work}
\subsection{MLOps and AutoML}  
MLOps, or Machine Learning Operations, is a way to manage and improve the process of building, deploying, and maintaining machine learning models. It brings together data scientists and operations teams to ensure that machine learning models can be used in real-life applications, just like regular software. MLOps helps track data, train models, and put them into production faster and more efficiently, allowing businesses to use AI in their day-to-day work~\cite{kreuzberger_machine_2023}.
Integrating ML workflows with traditional software pipelines is a key challenge in MLOps. Unlike regular software, ML involves data preprocessing, model training, and evaluation, requiring specialized tools and environments. These processes do not fit easily into standard DevOps frameworks, making automation and deployment difficult~\cite {breck_ml_2017}.

AutoML has emerged as a pivotal subfield within MLOps, aiming to streamline and optimize the ML workflow by automating processes such as model selection, hyperparameter tuning, and end-to-end pipeline creation~\cite{jung-lin_lee_human---loop_2019}. 
While AutoML has democratized ML by allowing non-experts to build sophisticated models, it also faces challenges in usability and adaptability. Researchers highlight that the complete automation in existing AutoML systems may be insufficient in complex or high-stake scenarios, where user expertise and interpretability are critical~\cite{jung-lin_lee_human---loop_2019}. 

Furthermore, the current landscape of AutoML tools is diverse, encompassing open-source libraries, cloud-based platforms, and enterprise-level solutions, each providing varying levels of support across the ML pipeline. Open-source tools such as auto-sklearn~\cite{auto_sklearn} and TPOT~\cite{tpot} have focused on specific tasks like model selection and hyperparameter tuning, while cloud solutions (e.g., Google Cloud AutoML~\cite{google_automl} and enterprise platforms (e.g., DataRobot~\cite{datarobot}) often offer a more comprehensive approach. This variation impacts user experience, particularly for practitioners who require seamless integration across data preprocessing, model training, and post-processing stages~\cite{xin_whither_2021}. Despite the benefits of these diverse tools, adoption remains relatively low in practice; empirical studies indicate that fewer than 2\%\ of ML practitioners on platforms like OpenML use AutoML, which is partly due to usability and a lack of transparency~\cite{majidi_empirical_2022}.



\subsection{Continuous experimentation}
\label{sec:related-cont-exp}
Continuous controlled experimentation is the practice of evaluating the impact of changes (e.g., addition or removal of features) on end users~\cite{kohavi_trustworthy_2020}.
It typically relies on performing randomized controlled trials or A/B tests. 
As a key enabler for innovation via data-driven decisions, this practice is embraced by several web-facing companies~\cite{kevic_characterizing_2017,tang_overlapping_2010,diamantopoulos_engineering_2020}. 
To help software-intensive companies become data-driven, Fabijan et al. defined the 
\textit{Experimentation Evolution Model} (EEM), i.e., a process of moving from ad-hoc data analyses to structured and scalable continuous experimentation~\cite{fabijan_evolution_2017}.
EEM breaks down the evolution into four phases---``crawl'', ``walk'', ``run'', and ``fly''---representing the increasing sophistication of the experimentation approach.
In ``crawl'', the technical focus is on setting up systematic logging to collect user interaction data.
Experiments are manually conducted without a dedicated experimentation platform, and data scientists play a crucial role in guiding product teams through the setup and analysis of these initial experiments. 
In ``walk'', metrics are defined from collected signals, and a dedicated experimentation platform is used (either internally built or adopted). The platform includes key features like power analysis and A/A testing, which help automate and streamline the experimentation process. 
Product teams begin to take on more responsibility for creating experiments, although data scientists still handle much of the execution and analysis. 
By the ``run'' phase, experimentation becomes more pervasive, with product teams managing a higher volume of experiments using more abstract and comprehensive metrics. 
Automated alerting systems and experiment iteration support are introduced to optimize experiment workflows.
Finally, in the ``fly'' phase, experimentation is integrated into every product change, from major features to minor bug fixes.
Advanced features, such as real-time detection of harmful experiments and institutional memory (which keeps a history of all the experiments), prevent redundant experiments.
The organization in this phase has reached a point where continuous experimentation is embedded in all aspects of development, driving data-driven decision-making at scale.
\textit{Taking inspiration from EEM, our MLOps approach aims to integrate continuous experimentation into all the phases of ML development, streamlining the work of data scientists. 
}
\begin{figure}[tb]
    \centering
    \includegraphics[width=\columnwidth]{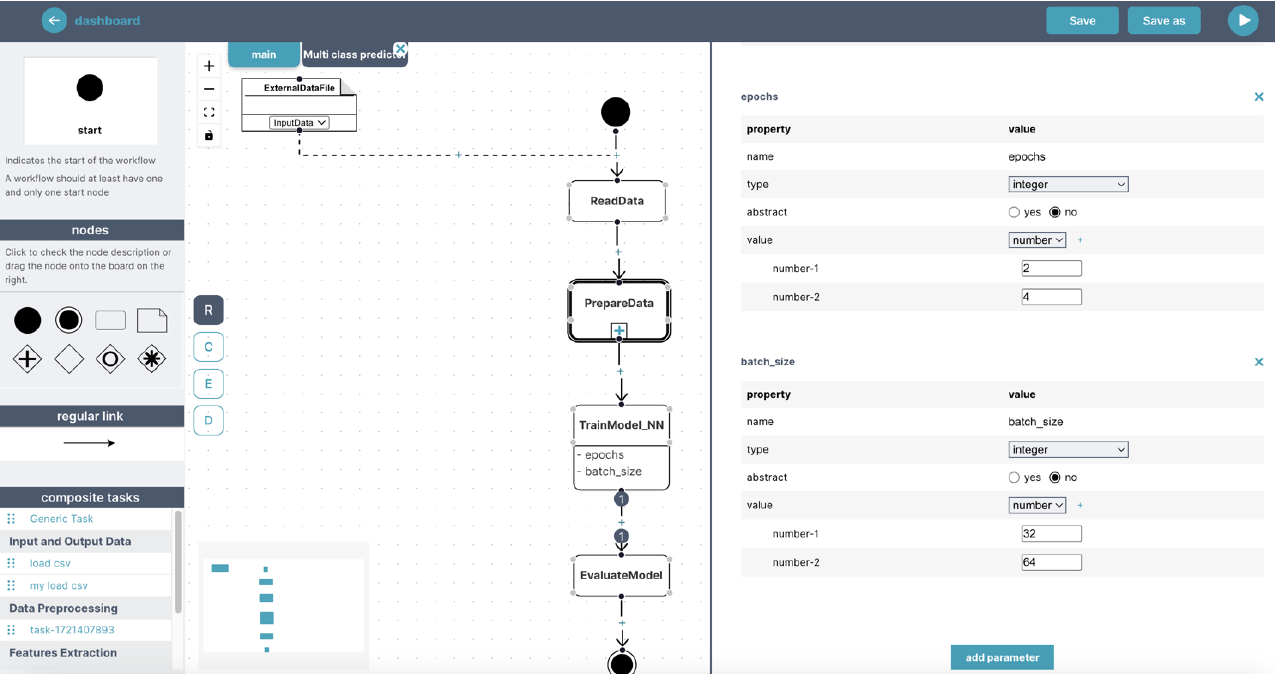}
    \caption{Experiment Workflow Designer Tool of \projectName{} portal.}
    \label{fig:Usecase-Implementation}
    \vspace{-1.0em}
\end{figure}

\subsection{Related work}
Several popular tools and frameworks support MLOps, including MLFlow~\cite{mlflow}, ZenML~\cite{zenml}, Neptune.ai~\cite{neptune_ai}, Kubeflow~\cite{kubeflow}, SageMaker~\cite{sagemaker}, and Weights \& Biases~\cite{weights-biases}. 
These tools certainly offer several features that can be used in the experiment-driven MLOps approach we advocate for. 
Namely, each of these tools supports traceability, maintains experiment history, and allows for replicating experiments. Most of them also provide insights into experimentation costs and trade-offs (e.g., resource consumption). 
Finally, all of them support some way of obtaining human feedback during an experiment. 

Our framework (\autoref{sec:extremexp-approach}) includes the above features, but also has several unique capabilities:

\begin{itemize}
    \item \textbf{Knowledge Repository}: It provides enhanced reusability, replicability, and knowledge alignment by having a unified semantic-graph-based repository for experiment specifications, evaluated workflows, user profiles, and derived knowledge from past experiments.
    \item \textbf{Complex Experimentation Strategies}: It facilitates complex experimentation strategies that take advantage of domain knowledge (e.g., dependencies between variability points in different tasks of a workflow).
    \item \textbf{Strategy Recommendation}: It supports data scientists by recommending efficient experimentation strategies based on the knowledge from past experiments. 
    
\end{itemize}

\section{Motivating Example}
\label{sec:running-example}


Imagine Andrea, a data scientist responsible for predictive maintenance in a large manufacturing plant. Andrea’s task is to forecast equipment failures using sensor data streamed from various machines. To achieve this, they set up a workflow composed of the tasks of \textit{Read Data, Add Padding, Split Data, Train Model, and Evaluate Model}. Within this workflow, the \textit{Train Model} task serves as an abstract element, accommodating different ML model types like standard, recurrent, and convolutional neural networks, allowing Andrea to experiment with different ML algorithms.

For Andrea, effective maintenance prediction is critical---not only for minimizing downtime but also for optimizing operational costs. However, achieving the highest accuracy requires more than just crafting a single model; it needs a systematic process of experimentation. Andrea must evaluate how factors like dataset granularity, model hyperparameters, and different ML algorithms and architectures influence the predictions. This experimentation process allows them to select ML models that maximize predictive accuracy without exceeding infrastructure and user constraints, such as resource (e.g., memory) consumption, and inference latency.

The need for flexibility in experimentation extends to both task-level and parametric variability. 
For instance, Andrea can interchange data processing steps, swap in alternative model training techniques, or adjust parameters like training thresholds. 
This way, an initial workflow can be dynamically configured to match a particular experimentation scenario.
By carefully evaluating different workflow configurations, Andrea gains insight into which of them works best under specific infrastructure and user constraints.

Over time, Andrea gains a deeper understanding of the factors that impact prediction accuracy and resource efficiency. This knowledge enables them to make better decisions, continuously enhancing the workflow's effectiveness in meeting the maintenance goals. 
Ideally, Andrea would like to follow a tool-supported experimentation process where knowledge of past experiments, performed by themselves or other data scientists, is used to tune the overall process---and they are involved only when and where they are really needed. 

\section{\projectName{} Framework}
\label{sec:extremexp-approach}


To support data scientists like Andrea, in \projectName{}, we focus on providing a complete technical framework for continuous improvement of ML development, maintenance, and evolution.


The framework's cornerstone is a web portal containing the \textit{Experiment Workflow Designer Tool}, which allows data scientists to prepare workflows either as textual specifications or visually (\autoref{fig:Usecase-Implementation}).
The tool provides a library of reusable tasks and (sub)workflows, from which workflows can be quickly composed---and newly designed ones can be stored there for further reuse.
MLOps experiments are also defined in the tool by defining associated workflows and experiment strategies (configurations to be evaluated, metrics to be computed). 
%
The framework provides an \textit{Experimentation Engine}, which takes the definition of an experiment and, based on the specified experiment strategy, executes the experiment automatically or semi-automatically (with input from the user). 
The heart of the platform is the \textit{Knowledge Repository}, which stores all results (data, metrics, etc.) from executing experiments as well as the definition of the experiments themselves.

The designer tool, the experimentation engine, and the associated metamodels are open-source projects under ongoing development\footnote{\url{https://github.com/ExtremeXP-VU/ExtremeXP-portal}, \url{https://github.com/ExtremeXP-VU/ExtremeXP-experimentation-engine}
\url{https://github.com/ExtremeXP-VU/extremexp-workflow-metamodel}}. 
In the rest of the section, we provide an overview of the main concepts and highlight the key innovations behind our work so far.  

\subsection{Main concepts and architecture}
\label{sec:extremexp-overview}






\begin{figure}
    \centering
    \includegraphics[width=\columnwidth]{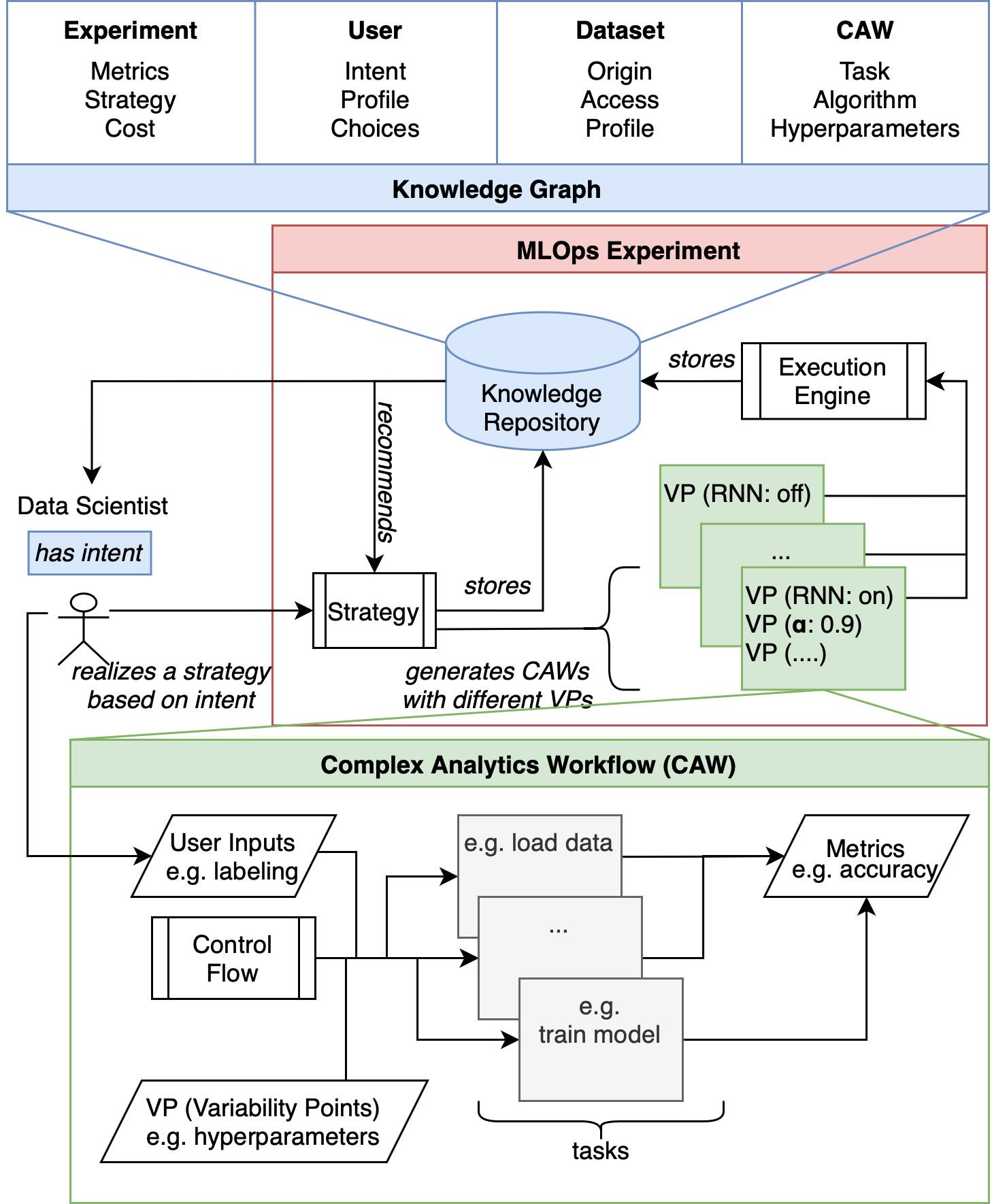}
    \caption{Main concepts of \projectName{} framework.}    
    \label{fig:ex-caw}
\end{figure}

In \projectName{}, we employ two main concepts: Experiment and Complex Analytics Workflow (CAW).
An \textbf{Experiment} is a structured user-centered optimization process with the aim to, e.g., create an optimized ML model based on the data and the knowledge gathered during similar past experiments.
An experiment consists of multiple CAWs. 
The experiment uses the past knowledge to execute CAWs to train different variants of the model and to optimize the model---while at the same time building knowledge used in future experiments.
The experiment is always built with a specific experimentation \textit{intent} and is controlled by an experimentation \textit{strategy}.

As depicted in \autoref{fig:ex-caw}, the experimentation intent coming from the data scientist (e.g., \textit{``find classifier with highest accuracy''}) is transformed into an experimentation strategy that prescribes which CAWs to run. 
This can be calculated once at the start (when using, e.g., full factorial design) or dynamically determined during the experiment execution (when using, e.g., bayesian optimization).
The decisions on how to steer the experiment and what CAWs to execute also take into consideration the experimentation \textit{cost}, i.e., the resources consumed for executing an experiment.
Such cost can include properties ranging from purely technical (e.g., memory, CPU time, energy) to high-level ones (e.g., overall execution time, attention span of data scientists), and can be estimated from past experiments. 

A \textbf{CAW} in \projectName{} is an abstraction that generalizes ML training/serving workflows and encompasses other workflow types such as simulation-based and data analytics-based workflows. 
CAWs consist of a number of tasks (e.g., `load data', `preprocess', `train', `evaluate') coordinated by a control flow. 
Importantly, tasks can be either \textit{automated} or \textit{manual}, i.e., tasks that require user input (e.g., labeling of a result/action). 

A CAW is generated by choosing a value for each \textit{Variability Point} (VP) of the experiment.  
VPs include (i) different task implementations (e.g., different ML algorithms), (ii) different task inputs, (iii) different hyperparameter values, and (iv) different task/workflow deployments (on CPUs vs GPUs). 

The execution of a CAW produces several \textit{metrics}. These are measurable properties of (i) the whole CAW (e.g., end-to-end execution time), (ii) a particular task (e.g., memory consumption of `train'), or (iii) an output of a task (e.g., accuracy of produced ML model or user satisfaction level given a task’s output).

  
\subsection{Learning from the past}
\label{subsec:learning-from-past} 

A complete MLOps experiment generates large amounts of data, which are not only about the results of the CAWs but, among others, they also include data about the experiment definitions and about user interactions with the framework. These data are useful for continuous improvement, as they can be leveraged to create experiments that are tailored to the data and the users’ needs. Therefore, in our framework, this information is stored in a Knowledge Repository, which is modeled in the form of a Knowledge Graph (KG). 

In the past, different KGs or ontologies have been defined to capture the different components of ML experiments~\cite{dmop,bigowl}, but none of them focuses on capturing the user for whom the experiments have been generated. To this end, our KG contains classes that are able to capture the general ML experiment components, such as algorithms, hyperparameters, metrics, or datasets, but also user-related concepts, such as their characteristics (e.g., domain of expertise, ML proficiency, etc.), their experimentation intent, their hard and soft constraints, their interaction with the system and their feedback. 

    
The stored knowledge is used to support users when designing new experiments by providing intelligent recommendations that are both context-aware and flexible~\cite{pons2024capturinganticipatinguserintents}. 
To achieve context-awareness, we rely on \emph{Knowledge Graph Embeddings}, which capture relationships and similarities within the data, giving a broader context to each entity captured in the KG. 
To achieve flexibility, \emph{Link Prediction} techniques have been adopted, allowing the system to predict potential missing links between entities. 

For illustration, when Andrea initiates a new experiment in the motivating example, a node is created to represent it. To generate its embedding, past-learned information from Andrea, their intents, and the datasets used, among others, are leveraged through their embedding representations. Link Prediction is then applied, based on the embedded information from past experiments, to identify the best matches for the elements to be recommended, such as the algorithms (i.e., inside the tasks of a CAW) to use or the metrics to assess. These embeddings are designed to evolve over time based on data from new experiments.

\subsection{Keeping human in control}

In our framework, the user maintains control over the experimentation process, with the ability to intervene at key interaction points. These points serve distinct roles:

\begin{enumerate}
    \item \textbf{User as supervisor:} Users can oversee the experiment’s progress, reviewing samples or intermediate results as desired. This supervisory role helps ensure that outcomes align with user expectations (user intent).
    \item \textbf{User as validator:} Users can act as validators, offering feedback on the accuracy of predictions or model outputs (e.g., ``Are these predictions accurate?'').
\end{enumerate}

These interaction points are flexible and skippable, allowing users to engage at critical stages such as reviewing intermediate results or bypassing these steps when desired. 
This flexibility allows users to intervene only when necessary, focusing their involvement on key decision points. 
Over time, the experimentation engine automates user validation by building \textit{user profiles} based on past interactions with the system. 
This allows for a more streamlined, automated optimization process that does not burden the data scientist while, at the same time, does not compromise quality.

In particular, to decide when certain interactions can be skipped, we rely on (i) the availability (and eventually also the quality and trustworthiness) of user profiles and (ii) the \textit{interaction budgets} provided by users. 
An interaction budget specifies the amount of interaction (in terms of, e.g., minutes) that is acceptable by a user. 
This amount of interaction increases by involving the user and decreases by skipping interactions. 
The degree of increase also depends on the mental effort expected by the user, or, in other words, the difficulty of the task.

\subsection{Continuously improving} 
\label{sec:continuous-improving}

%


Continuous improvement of ML models and of their development and evolution is enabled in our framework by (i) maintaining an institutional memory of past experiments, and (ii) keeping experiments active even after the ML models optimized by them have been deployed to production. 

The first idea is directly implemented via the Knowledge Repository (KR) (\autoref{subsec:learning-from-past}).
This contains all the history of executed experiments, as well as the traceability between such experiments and generated data, models, and users.  
The larger the KR grows, the more support the framework can provide in the form of, e.g., preventing users from running redundant or highly counterproductive experiments. 
Note that a \textit{productive} experiment for \projectName{} is one that (i) directly contributes to satisfying a user's intent (e.g., \textit{``find best classifier''}) and/or (ii) contributes to the institutional memory of the framework, and hence to its continuous improvement.





The second idea can be illustrated in the motivating example (\autoref{sec:running-example}). 
The result of the experiment in the example is a trained predictor.
However, our approach does not stop when the experiment is finished.
As the predictor is deployed in production, its performance and efficiency are monitored, and these metrics are also stored in the KR.
If there is enough new data or the performance of the predictor deteriorates (e.g., due to distribution shift), the experiment can be re-executed with the newly collected data. 
The KR serves here to speed up the experiment by allowing it to omit training of model variants which are likely to have poor performance. 
It is also possible to plan the cost of the retraining by having data regarding the cost from the previous experiments.
The original experiment is thus extended to a series of experiments that can run for as long as the system is deployed.
With this approach, we are moving even beyond the EEM's ``fly'' phase (\autoref{sec:related-cont-exp}) as we actively use results from previous experiments to re-run them and, potentially, obtain better results.

\section{Applications and Experience}

The \projectName{} project has five real-life use cases that span the diverse domains of transportation, emergency coordination, flash flooding, industrial manufacturing, and cyber-security.
We have so far been successful in modeling and executing at least one experiment for each use case using our domain-specific textual and graphical languages and tools. 
Having a small set of well-defined concepts and dedicated tools has dramatically improved the communication between domain experts (e.g., in transportation/flooding simulations) and data scientists in the project. 
Our definition of CAW as a workflow that in principle contains not only automated but also manual tasks aligns well with the way most data scientists in these diverse domains approach problems, namely as a combination of manual and automated work. 
We have also observed that all of them value the ability to replicate experiments in a trustworthy way, and, even more importantly, having a system that can reason on top of the generated knowledge and help them in taking decisions tailored to their needs (or ``intents'').
Finally, all of them stressed the need for being able to interact with a running experiment in an efficient and effective way, e.g. by being presented with an overview of the CAWs executed so far and zooming in only when needed. 

Given the above experience so far, we believe that the new paradigm of continuous experiment-driven MLOps embodied in the \projectName{} framework holds the potential for supporting data scientists in real-life ML-enabled systems. 
As immediate next steps, we are working on further developing the concepts of interaction budget and experimentation cost and their interplay with even more flexible experimentation strategies.

\section*{Acknowledgment}
This work has been partially supported by the EU project ExtremeXP grant agreement 101093164, by Charles University institutional funding SVV 260698, and by the Spanish Ministerio de Ciencia e Innovación under project PID2020-117191RB-I00/ AEI/10.13039/501100011033 (DOGO4ML).
\bibliographystyle{IEEEtran}
\bibliography{paper_cain}

\end{document}